\documentstyle[epsfig,mathptm]{mn}
\DeclareMathVersion{bold}
\newif\ifAMStwofonts
\AMStwofontstrue                % <- comment this out if AMS fonts not available

\def\deg{\nobreak^\circ}

\DeclareMathVersion{bold}                    % <- seems to be needed with mathptm
\DeclareMathAlphabet{\mathbfit}{OT1}{cmr}{bx}{it}
\SetMathAlphabet\mathbfit{bold}{OT1}{cmr}{bx}{it}
\DeclareMathAlphabet{\mathbfss}{OT1}{cmss}{bx}{n}
\SetMathAlphabet\mathbfss{bold}{OT1}{cmss}{bx}{n}
\ifAMStwofonts
  \ifCUPmtlplainloaded \else
    \DeclareSymbolFont{UPM}{U}{eur}{m}{n}
    \SetSymbolFont{UPM}{bold}{U}{eur}{b}{n}
    \DeclareSymbolFont{AMSa}{U}{msa}{m}{n}
    \DeclareMathSymbol{\upi}{0}{UPM}{"19}
    \DeclareMathSymbol{\umu}{0}{UPM}{"16}
    \DeclareMathSymbol{\upartial}{0}{UPM}{"40}
    \DeclareMathSymbol{\leqslant}{3}{AMSa}{"36}
    \DeclareMathSymbol{\geqslant}{3}{AMSa}{"3E}
  \fi
\fi
\title[Galactic emission in Tenerife data]{On dust-correlated 
Galactic emission in the Tenerife data}
\author[P. Mukherjee et al.]
{P. Mukherjee, A.W. Jones, R. Kneissl \& A.N. Lasenby\\
Astrophysics Group, Cavendish Laboratory, Madingley Road, 
Cambridge CB3 OHE, UK\\}
\date{Accepted    . Received    ; in original form 15th of February 2000}
\pagerange{\pageref{firstpage}--\pageref{lastpage}}
\pubyear{2000} 

\def\LaTeX{L\kern-.36em\raise.3ex\hbox{a}\kern-.15em
    T\kern-.1667em\lower.7ex\hbox{E}\kern-.125emX}

\begin{document} 

\maketitle

\label{firstpage}

\begin{abstract}
Recently correlation analyses between Galactic dust emission templates 
and a number of CMB data sets have led to differing claims on the origin 
of the Galactic contamination at low frequencies.  de Oliviera-Costa 
{\em et al} (1999) have presented work based on Tenerife data supporting the 
spinning dust hypothesis.  Since the frequency coverage of these data is 
ideal to discriminate spectrally between spinning dust and free-free 
emission, we used the latest version of the Tenerife data, which have 
lower systematic uncertainty, to study the correlation in more detail.  We 
found however that the evidence in favor of spinning dust originates 
from a small region at low Galactic latitude where the significance of 
the correlation itself is low and is compromised by systematic 
effects in the Galactic plane signal.  The rest of the region was found to be 
uncorrelated.  Regions that correlate with higher significance 
tend to have a steeper spectrum, as is expected for free-free 
emission.  Averaging over all correlated regions yields dust-correlation
 coefficients of $180\pm47$ and $123\pm16$ $\mu$K /MJy sr$^{-1}$ at 10 and 
15~GHz respectively.
  These numbers however have large systematic uncertainties that we have
 identified and care should be taken when comparing with results from other
 experiments.  We do find evidence for 
synchrotron emission with spectral index steepening from radio to microwave 
frequencies, but we cannot make conclusive claims 
about the origin of the dust-correlated component based on the spectral 
index estimates.  Data with higher sensitivity 
are required to decide about the significance of the dust-correlation 
at high Galactic latitudes and other Galactic templates, in particular 
H$_\alpha$ maps, will be necessary for constraining its origin. 
\end{abstract}

\section{Introduction}
\label{intro}
A small, but significant correlation of existing Cosmic Microwave
Background (CMB) data with maps of Galactic dust emission has been 
detected. It is important to understand the origin, and hence the
characteristics of any Galactic emission present in CMB maps with
structure on angular scales relevant to CMB measurements.
With such information, we can attempt to remove these contaminating
emissions from the data, to be left with a pure CMB map. 

Cross-correlating the COBE DMR maps with
DIRBE far-infrared maps, Kogut {\em et al} (1996a,b) discovered that 
statistically significant correlations did exist at each DMR frequency, 
but that the frequency dependence was inconsistent with vibrational dust 
emission alone and strongly suggestive of additional dust-correlated 
free-free emission ($\beta_{ff} = -2.15$).  A number of recent microwave 
observations have also shown that these 
correlations are not strongly dependent on angular scale. 
Different experiments ( DMR: Kogut et al. 1996b; 19GHz: de
Oliviera-Costa et al. 1998; Saskatoon: de Oliviera-Costa et al. 1997;
MAX5: Lim et al. 1996; OVRO: Leich et al. 1997), together give, with
$95\%$ confidence, $-3.6 < 
\beta_{radio} < -1.3$, consistent with free-free emission over the
frequency range 15-50~GHz.  However if the source of this correlated
emission is indeed free-free, it is expected that a similar
correlation should exist between $H_\alpha$ and dust maps.  Several
authors have found that these data sets are only marginally correlated
(McCallough 1997 and Kogut 1997).  Thus, most of the correlated
emission appears to come from another source. 

Draine and Lazarian (1998a,b) suggest that the correlated emission could
originate from spinning dust grains.  This model predicts a microwave
emission spectrum that peaks at low microwave frequencies, the exact 
location of the peak depending 
on the size distribution of dust grains, 
and has a spectral index between
-3.3 and -4, over the frequency range 15-50~GHz (Draine and Lazarian 1998b, 
Kogut 1999).
  The spectral indices
obtained from various experiments do not seem to conform too well to
the spinning dust model alone. However, recently de Oliveira-Costa
{\em et al}
(1999) (hereafter DOC99) claim to have found evidence of the spinning
dust origin of DIRBE-correlated emission using the Tenerife 10 and
15GHz data.  They find a rising spectrum from 10GHz to
15GHz, which is indicative of a spinning dust origin, as opposed to a
free-free origin, for the dust-correlated Galactic foreground at these
frequencies.  In this paper, we use the most recent Tenerife data to show
that the correlated emission is not necessarily indicative of spinning
dust. 

\section{The data and Galactic templates}

Tenerife is a double differencing experiment that takes data by drift
scanning in right-ascension at each declination.  Data are taken every 
$1\deg$ in RA 
and at intervals of $2.5\deg$ in Dec.  We used the Tenerife
10GHz and 15GHz data in the region, $32.5\deg < $Dec$ < 42.5\deg$, 
and $0\deg < $RA$ <360\deg$, with point sources subtracted and 
baselines removed.  The Tenerife experiment has beams of 
FWHM $4.9\deg$ and $5.2\deg$ for the 10 and 15GHz experiments 
respectively, with a differencing angle of $8.1\deg$ between east and 
west positions on the sky to make up the final triple beam (Jones {\em et al}
 (1998)).

Figure 1 shows the patch of sky covered by the experiment in Galactic 
projection.  Figure 2 shows the dust data and the Tenerife data at 
all the five declinations where data have been taken. 
We have extended the 
analysis presented in Gutierrez {\em et al} (1999) to also cover a region 
outside the high latitude range ($b>40\deg$) that they use to 
constrain CMB
fluctuations. Therefore, the data we have used are different from 
that used in DOC99 as they did not extend the coverage of the raw data
processing to be outside this range.  This is one of the reasons we would 
expect 
differences between our results and those of DOC99. A great advantage of 
using the Tenerife data to constrain the spectral index of the 
dust-correlated component lies in the fact that similar angular scales 
and the same patch of the sky can be used. This is not the case when the 
correlation from different experiments are compared. Therefore, we have 
not used the two further declinations covered by the 15GHz data set 
(Decs $30\deg$ and $45\deg$), but not by the 10GHz data, avoiding any 
effect from features present in these declination strips. However, 
we have redone the calculations using these extra declinations and there 
was no significant difference with the results presented in this paper.  

When looking at Figure 2, a difference between the two sides of the 
RA range, at almost all declinations is apparent. 
The rise (drop because of the double differencing) of the 
Galactic signal towards low Galactic latitude seen in the 
convolved dust maps on both sides of the RA range is 
only followed on one side by the 
Tenerife data. The Tenerife Galactic signal shows a stronger Galactic 
centre / anti-centre difference and a variation with declination which
is different compared to the dust template with a particularly strong 
feature at Dec $42.5\deg$. 
Also note that the noisier 10GHz 
data show a wider spread toward low Galactic latitude, which does not 
appear systematically Galactic, at least when compared to the Galactic 
dust template. This may be due to a systematic error in the Tenerife
data that is not fully represented by the error bars. 

The Galactic templates used are the destriped DIRBE+IRAS $100\mu$m 
template of 
dust emission (Schlegel, Finkbeiner and Davis 1998), the 408MHz 
survey of synchrotron emission (Haslam 
et. al. 1981), and the 1420MHz survey of synchrotron emission (Reich 
$\&$ Reich, 1988, hereafter $R \& R$). We used the cleaned and 
destriped versions of the synchrotron emission maps by Lawson, 
Mayer, Osborne \& Parkinson (1987). 

\begin{figure}
\centerline{\epsfig{
file=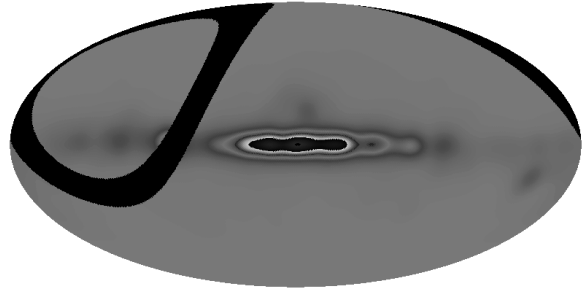,width=7.7cm,angle=0}}
\caption{The Figure shows the patch of sky covered by the Tenerife
experiment in black, overlayed on the DIRBE template (smoothed to
$5\deg$) to show the position of the Galactic plane.}
 \label{fig:plot1}
\end{figure}

\begin{figure}
\centerline{\epsfig{
file=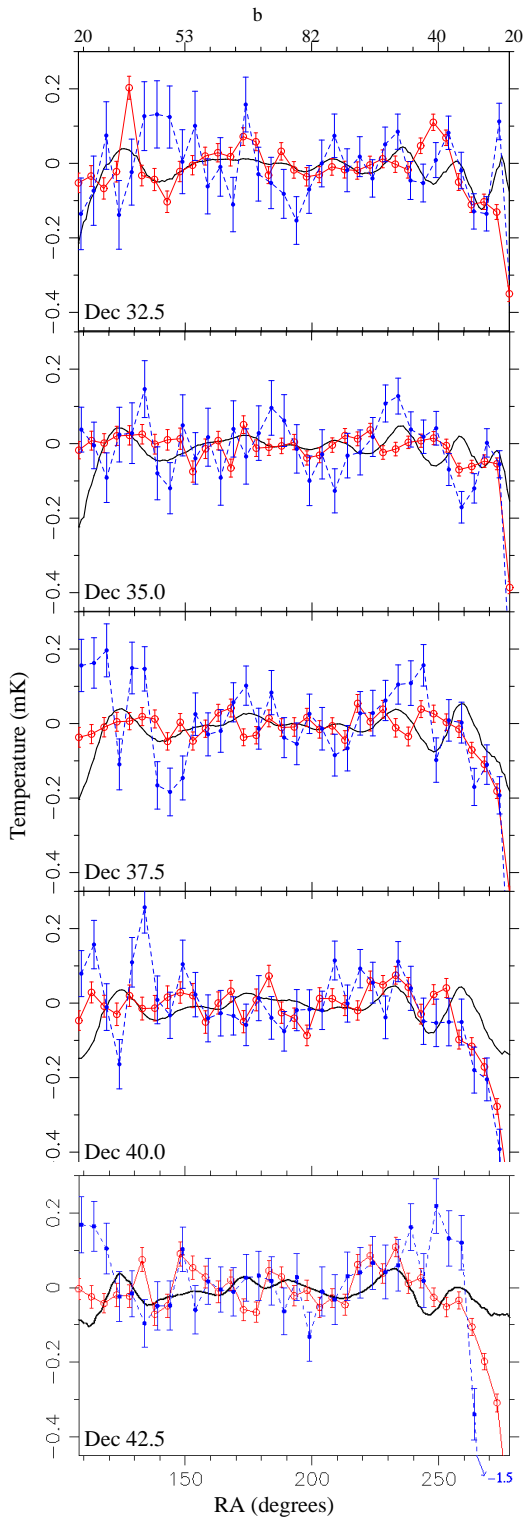,width=7.cm,angle=0}}
%file=alldata.eps,width=8.5cm,height=10cm,angle=0}}
\caption{The figure shows the Tenerife and dust data in the RA 
range that lies above the Galactic cut 
of $b>20\deg$, for all five declination strips.  The filled 
circles and the dashed line represent the Tenerife 10GHz data, 
the empty circles and thin solid line represent the 15GHz 
data.  The data have been binned every 5 points in RA.  The 
corresponding $1\sigma$ error bars are shown.  Units are mK.  The thick solid 
line is the $100\mu m$ data convolved with the Tenerife
beam.  This data has been plotted in units of 
$0.13\times 1$ MJy/sr.}
\label{fig:plot2} 
\end{figure}

\section{Correlation analysis and results}
\subsection{The method}

Assuming that the microwave data consist of a super-position of CMB,
noise and Galactic components, we can write 
$$y = aX + x_{CMB} + n.$$ 
Here $y$ is a data vector of $N$ pixels, $X$ is an $N\times M$ 
element matrix 
containing $M$ foreground templates convolved with the 
Tenerife beam, $a$ is a vector of size $M$ that represents 
the levels at which 
these foreground templates are present in the Tenerife data,
$n$ is the noise in the data and $x_{CMB}$ is due to the CMB 
convolved with the
beam. The noise and CMB are treated as uncorrelated. 

The minimum variance estimate of a is
$$ \hat{a} = [X^{T} C^{-1} X]^{-1} X^{T} C^{-1} y,$$
with errors given by $\Delta \hat{a}_i = \Sigma_{ii}^{1/2}$ 
where $\Sigma$ is given as,
$$ \Sigma = <\hat{a}^2> - <\hat{a}>^2 = [X^{T} C^{-1} X]^{-1}. $$
$C$ above is the total covariance matrix (the sum of the noise
covariance matrix and the CMB covariance matrix).  The noise
covariance matrix of the Tenerife data is taken to be diagonal, and
the CMB covariance matrix is obtained through Monte-Carlo simulations
with 10000 CMB realizations (this can of course be found analytically
as well, as in DOC99). In all the above equations
$X$ and $y$ are actually deviations of the corresponding quantities
from the weighted mean with weights\footnote{\small{We have tried 
a variety
of different weighting schemes, including uniform
weights, to check our method and all results found were within the
$68\%$ confidence limits of those presented in the paper.}}
given by $C^{-1}$ (this is 
different to DOC99 who use the actual $X$ and $y$ in the analysis).  
It should be noted that there may be other components of emission 
present in the data so that 
the errors we consider here are only lower limits. Also, the templates
that we are using may not be perfect and we do not account for
possible uncorrelated structure in $C$. If there are other
non-Gaussian components present in the data then the minimum variance
estimate of $a$ will be incorrect.

\subsection{Results}

Listed in Table 1 are the correlation coefficients obtained for
the Tenerife data sets and the $100\mu m$ DIRBE template, the Haslam,
and the $R\& R$, with each Galactic template taken individually.
Table 2 shows the correlation coefficients obtained when fits
are done for two templates jointly.  The DIRBE and Haslam correlations
correspond to joint $100\mu m$-Haslam fits, whereas the $R\&R$ values
correspond to joint $100\mu m-R\&R$ fits.  $\Delta T= (\hat{a}\pm \Delta
\hat{a})\sigma_{Gal}$ is the amplitude of fluctuations in temperature in
the Tenerife data, that results from the correlation ($\sigma_{Gal}$
being the $rms$ deviation of the template map).  The analysis is done
for different Galactic cuts.  Only positive Galactic latitudes are
used in these tables as the Tenerife data do not extend below
a $b$ of $-32\deg$. 

From Table 1, we see that correlations with each of the three
templates are detected with high significance.  However these
correlations fall off in significance towards higher Galactic
latitudes.  For comparison, the {\em rms} of the 10GHz data (after
subtracting noise) for $b > 20\deg$ is $181\pm 9
\mu K$ (the error is due to a 
$5\%$ calibration uncertainty). If the templates are not significantly
correlated with each other, one would conclude that 40\%
of the signal at 10~GHz is Galactic emission from Table 1. 
It should be noted that the correlation between the DIRBE and Haslam
Galactic templates is significant for all Galactic cuts although if the
dust and free-free emissions (or spinning dust) are 100\% correlated then
the Galactic emission would still only account for 50\% of the total
emission (assuming that all of the Galactic foregrounds present at this
frequency are present in either the Haslam or DIRBE templates).

We also see that the DIRBE-correlated component has slightly higher 
significance
than the Haslam or $R\&R$-correlated components  near the Galactic
plane for both the data sets. Away from the Galactic plane, at 10~GHz,
the component correlated to the Haslam map becomes dominant 
(for $b > 20\deg$) whereas at 15~GHz
the DIRBE component remains dominant.  However, there is almost no significant
correlated emission beyond this Galactic cut.  
 
\subsection{Discussion}

It is seen from Table 1 that individual correlations with dust
do not show a rising (spinning dust) spectrum with significance
between 10 and 15~GHz for any Galactic cut. From the joint $100\mu m -
Haslam$ correlations (Table 2), we see that it appears as if there is 
some evidence for a rising spectrum of dust-correlated emission only
for the $b > 20\deg$ cut (as reported by 
DOC99). Spectral indices inferred from the correlation coefficients are listed
in Table 3.  Errors are small close to the Galactic plane.
However, the values do not consistently agree with the spectral 
indices of any one of the three Galactic emission components,
 -2.8 for synchrotron, -2.1 for
free-free and positive for spinning dust, for example +1 
if the peak is at 20 GHz, that could be expected to be present 
as contaminating emissions in CMB data. 

From Table 3 we see that the value of the spectral index inferred from
the DIRBE-correlated component at 10 and 15~GHz is negative for most
Galactic cuts (the correlations for $b>30\deg$ and $b>40\deg$ are not
significant, so it was not possible to infer a spectral index\footnote{\small{
No significant correlations were obtained even for the $b>25\deg$ cut, 
indicating that only a small region near $b$~$20\deg$ Galactic latitude is 
correlated.}}).
From the positive value of the spectral index for $b > 20\deg$ DOC99
conclude that spinning dust is responsible for producing the major
part of the DIRBE-correlated emission.  However, since the 
spectral indices inferred for all other Galactic cuts are negative, 
a specific model of the spatial distribution of spinning dust would 
be required to explain the data. On the other hand it could also 
be that in the $b$-range where the Galactic signal in the Tenerife 
data drops the estimation of the spectral index becomes more sensitive 
to systematic effects we have not accounted for. To decide this we 
focus in on the $b > 20\deg$ range and divide the data further. 
It should be noted that these spectral indices were found 
using the values for $\hat{a}$ and not
$\Delta T$ as the beam sizes between the 10 and 15GHz data sets are
slightly different giving rise to a lower expected $\Delta T$
(although the same relative level of contamination and hence $\hat{a}$) 
at 15~GHz than at 10~GHz which would systematically bias the values
obtained. 

When we examine the spectral indices obtained from the component 
correlated to the Haslam map at the two frequencies, we find that the
values from the individual analysis are less negative than 
expected for synchrotron emission (Table 3).
Joint correlation of the $100\mu$m dust and Haslam maps with
the data result in spectral indices that are steeper than expected
for synchrotron. This seems to imply that most of the emission in the 15GHz 
data that is
 correlated with the Haslam map is also correlated with the DIRBE maps (this 
is expected as the region that correlates lies close to the Galactic plane), 
leading to a much lower value of $\hat{a}$ for the synchrotron-correlated
 component than expected at 15GHz (and hence a much steeper spectrum).  
This may simply be because dust-correlated 
emission is more significant in the 15GHz data.  Similarily since the Haslam
-correlated emission is more significant at 10~GHz, the joint correlation 
method gives a lower value of $\hat{a}$ for the dust-correlated component and
 a correspondingly higher value for the synchrotron-correlated component.
  Being able to correct for this effect might therefore sort the systematic 
discrepancy.  However, since we see this discrepancy with the synchrotron 
spectral index we should be wary of the dust template correlation as well. 

If we now consider the DIRBE template to be an exact predictor for either
spinning dust or free-free emission (so that all the features present
in this template are present in the Tenerife data; as we are assuming
when performing the correlation analysis) then the level of
`contamination' $a$ will be independent of the region tested. This
means that each value of $a$ in Table 1 should be the same for
the 10 and 15GHz data respectively. We can therefore take the variance
of $a$ across the different regions analysed as a measure of the error
in this method, and find that weighted mean and rms 
at 10~GHz and 15~GHz are 157 (20) and 
122 (10) $\mu$K / (MJy sr$^{-1}$) respectively. This would correspond
to a spectral index of $-0.6^{+0.5}_{-0.5}$.

\subsection{Further study of the $b > 20\deg$ cut region}

We now divide the data into halves in RA, which gives two regions, 
one towards the Galactic centre and one towards the Galactic 
anticentre, at each declination. Repeating our analysis on these 
data we quantify the visual impression from Figure 2. 
Although the rise of the Galactic plane signal is equally strong in both 
regions of the dust template, we only find a significant correlation 
with the 10 and 15GHz data in the Galactic centre region, in spite of the fact 
that the noise errors in both regions are comparable.
The results of this analysis are listed in table 4.  
It is seen that of the five declination 
stripes in the centre region only the three 
at higher declination are significantly correlated. 
The anticentre regions are found to be uncorrelated or even 
significantly anticorrelated in two stripes, 
which again demonstrates that the structure is not correlated 
rather than that our sensitivity is reduced when dividing the data. 
Again if we take the weighted mean and rms 
in the value of $a$ between declinations, we get 
197 (66) and 157 (91) $\mu$K / (MJy sr$^{-1}$) 
for the centre regions at 10 and 15~GHz respectively, 
giving a spectral index of 
$-0.6^{+2.2}_{-2.8}$.  The variation between declinations 
is high and the spectral index thus obtained is consistent 
with both free-free and spinning dust models.

We perform yet another split of the data in an attempt to identify 
the region where the correlation mainly comes from.  It was found 
that the 20 pixels at lowest Galactic latitude of 
all declinations taken together (100 pixels) gave 
significant correlations, while the remaining pixels (750) gave 
no correlation (see table 5). Again the spectral index for 
dust-correlated emission is small and negative, while that 
for emission correlated to the Haslam map is steeper than expected. 
Since it is clear that the different 
declinations correlate differently, correlating them together is incorrect. 
Hence, shown also in table 5 are the values obtained for the 
same analysis on the 20 pixels at lowest Galactic latitude of each 
declination.  Even though the signal to noise is much lower here, 
it is clear that the Galactic plane signal correlates with a free-free 
like spectral index while the rest of the pixels shows a significant 
anti-correlation with dust at 10~GHz for some declinations.  The Galaxy 
correlates positively while the remaining regions correlate 
negatively or insignificantly.  It is clear from analysing 
such splits that the correlation coefficients obtained when taking 
all declinations together as in tables 1 \& 2 ($b>20\deg$) and in table 
4 \& 5, and 
which seem to indicate an almost flat or less negative spectral index 
between 10 and 15~GHz, are composed of regions that correlate 
with a negative spectral index and regions that do not correlate 
significantly at all. The anti-correlation, which 
occurs mostly in the 10GHz data, lowers the combined best fit $a$ 
value at that frequency, and hence raises the spectral index of 
dust-correlated emission. 
Note here that the high values of 'a' obtained need not to be in 
contradiction to the level of free-free allowed 
by H$_\alpha$ as the detections are all close to the Galactic plane, 
where the level of free-free is expected to be high.  
We still see a large variance between the three declinations 
which correlate significantly.  If we take the
weighted mean and rms of the highest three declinations
as indicative of a composite correlation, we get 313 (172) 
and 219 (83) $\mu$K / (MJy sr$^{-1}$) at 10~GHz and 15~GHz respectively,
 giving a spectral index of $-0.9^{+2.8}_{-2.2}$, again consistent
with both free-free and spinning dust.  Note that this value was obtained 
from the weighted average of three pairs of numbers shown in table 5, 
where each pair indicated a negative spectral
index. The spectral index obtained simultaneously for the component 
correlated to the Haslam map is consistent within 1~$\sigma$ with the 
synchrotron spectral index.  

Given that the dust templates correlate with the data only in a small 
part of the Tenerife patch above $b = 20\deg$ we further 
test, how physically meaningful the detection is. This is important 
because the spectral index of a component in this method is inferred 
by assuming that the difference in the correlation amplitude between the 
frequencies is solely due to the change of emissivity of a single 
physical component and is not due to a spurious change in the strength 
of the correlation. 

\section{Sky rotations}

Another way to test for systematic errors in the method is to 
take random samples from the Galactic templates themselves as 
"Monte Carlo" simulations. The advantage is that we can test 
against chance correlations with typical structure in the templates 
without having to understand the templates to a degree that would 
allow us to model this structure. Since the rise of the signal towards 
the Galactic plane clearly is a strong feature in the data and 
high Galactic latitude correlations are at best weak, we correlate 
to rotated / flipped maps in the northern and southern hemispheres, 
which all have the same orientation relative to the Galactic plane. 
Figure 3 shows $\hat{a}$ and its significance for 10 and 15~GHz, for 
the $b = 20\deg$ Galactic cut, the Galactic centre region of all declination
 strips taken together, 
and the spectral indices derived from these values. We find that the 
real patch does not correlate most significantly at either frequency. 
Given the number of patches that correlate more significantly we 
conclude that the real correlation is at most typical and only 
due to the Galactic plane signal. Further we see 
that the 15~GHz real correlation although less 
significant than the neighbouring points has a higher value of 
$\hat{a}$. It seems that the real dust patch, although it does not 
represent the observed 15~GHz structure particularly well, 
happens to produces a high correlation amplitude. Indeed the 
real dust patch shows no strong features (low intensity rms), 
which explains the weak correlation and results in an overestimate 
of $\hat{a}$ at 15~GHz. An interesting point is that the mean 
spectral index which could be interpreted as the spectral index 
derived from correlations to typical Galactic plane dust signals, 
correlations which are about as significant as the real correlation, 
has a value of $-1.5\pm 1.2$, well consistent with free-free emission.
The $rms$ variation in the value of the correlation coefficient at 
10 and 15~GHz can be taken as an estimate of the error in $\hat{a}$ for 
the real patch.  The error estimates obtained in this way range from about
 $20\%$ for $a$
 at both 10 and 15~GHz for the $b>0$ Galactic cut, to $64\%$ and $30\%$
 at 10 and 15~GHz respectively for the Galactic centre regions in the $b>20$ 
Galactic cut.  As expected, the errors from sky-rotation increase with 
decreasing signal and size of the region and are very large 
for the individual declinations.  Note that the lower Galactic cuts give the 
highest correlation with the highest significance for the real patch 
showing that the correlations there are not just due to aligned structure 
of the typical Galactic plane.  The error estimates presented in this section
are similar whether we include or exclude the rotated patches near the 
Galactic centre and anti-centre, where significantly different 
structure could be expected, though in the first case errors increase 
slightly for both 10 and 15~GHz.  Henceforth we use results corresponding to 
the second case. 

Since we have strong indication that the correlations between 
dust and the Tenerife data results only from the Galactic plane 
signal which will be similarly present in many 
physical components of Galactic emission we further investigate 
this particular feature of spatial distribution by modelling it 
in the next section. 

\begin{figure}
\centerline{\epsfig{
file=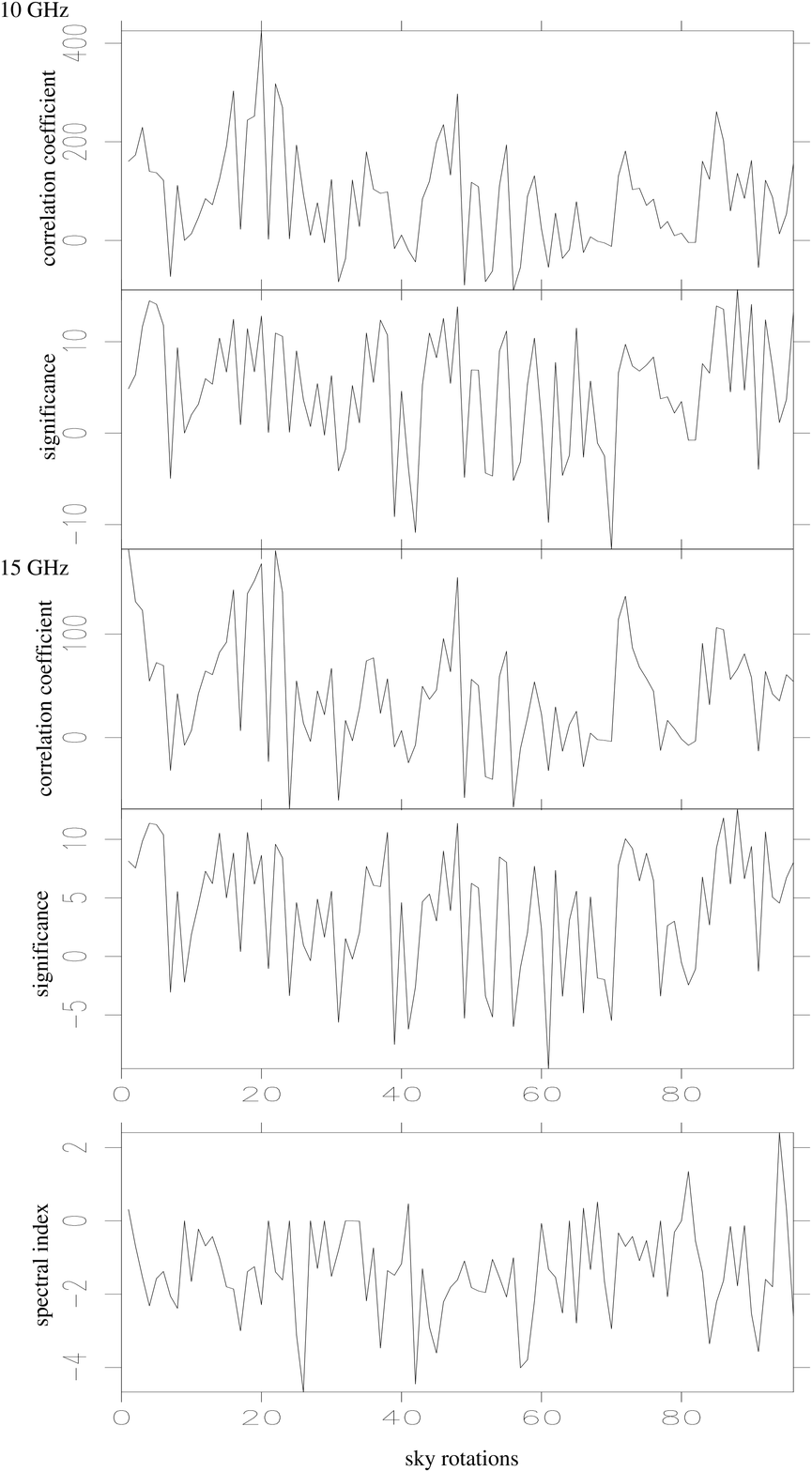,width=7cm,height=9cm,angle=0}}
\caption{The top four panels show the correlation coefficients and 
their significances when the 10 and 15~GHz data are correlated with 
different rotated patches of the dust template.  These are joint 
correlations of the centre regions of all declinations combined, when the 
Haslam map is held fixed. The bottom panel shows the corresponding 
spectral indices. The first point in each panel corresponds to the real 
patch.}
\end{figure}

\section{Galaxy modelling}

A problem with the interpretation of the correlation results
at different frequencies purely as a spectral dependence is that it
neglects the possibility of variations in the shape of the Galaxy
due to different components at different frequencies, and the resulting 
change of spatial alignment.  
For example, this method for obtaining the spectral index of correlated 
emission does not take into account the varying extent of the galaxy 
in the different data sets and template maps, so that the errors that 
we are quoting in the above tables might be too small and could be 
systematically wrong. If the Galaxy has a different spatial extent at 
10~GHz, 15~GHz and 3~THz ($100~\mu$m), it will result 
in a higher or lower value for the overall correlation coefficient
than is actually the case.  We investigate this effect by modelling
the spatial distribution of Galactic emission in the next section.  

\begin{figure}
\centerline{\epsfig{file=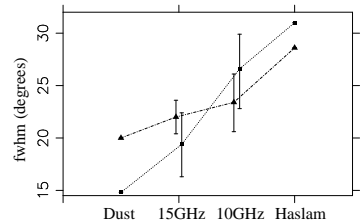,height=3cm,angle=0}}
\caption{The Figure shows the FWHM of the Gaussian model for the Galaxy 
that gave the most significant correlation (or minimum $\chi^2$) plotted 
against the different data sets for the $b > 20\deg$ Galactic cut and 
for Dec 40.0 (triangles) and Dec 37.5 (squares). 
The model was correlated with the Galactic centre regions of the data only.}
\end{figure}

\begin{figure*}
\centerline{\epsfig{
file=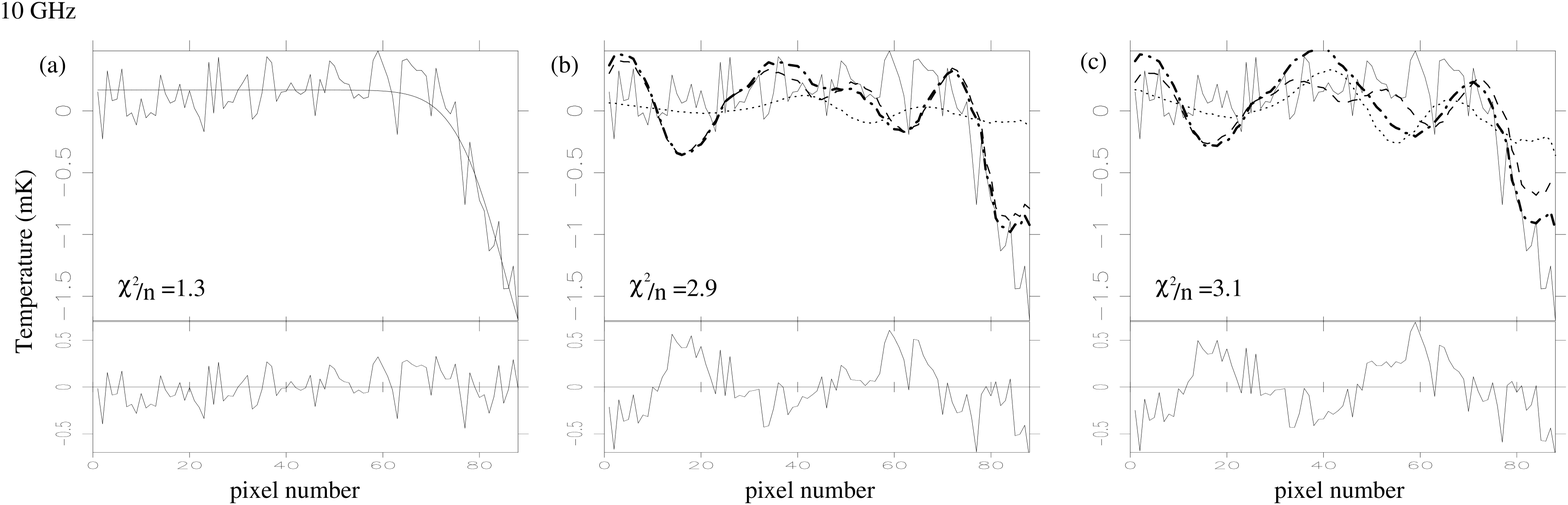,height=4.5cm,width=15cm,angle=0}}
\caption{Fitting the structure in the Tenerife data with the Gaussian model 
and the Galactic templates. Examples are shown for the 10 GHz data  
at Dec 42.5: (a) shows the best fit Gaussian model to the 
data and residuals 
(b) shows the Haslam (dashed) and dust (dotted) templates with amplitudes 
from the joint fit result and the thick dot-dashed line is the 
sum of the two 
(c) an equally acceptable fit, given the goodness of fit, is achieved 
when the spectra are normalised with the amplitudes of the templates 
at 15 GHz and are constrained to free-free and synchrotron spectral 
indices.} 
\end{figure*}
\begin{figure*}
\centerline{\epsfig{
file=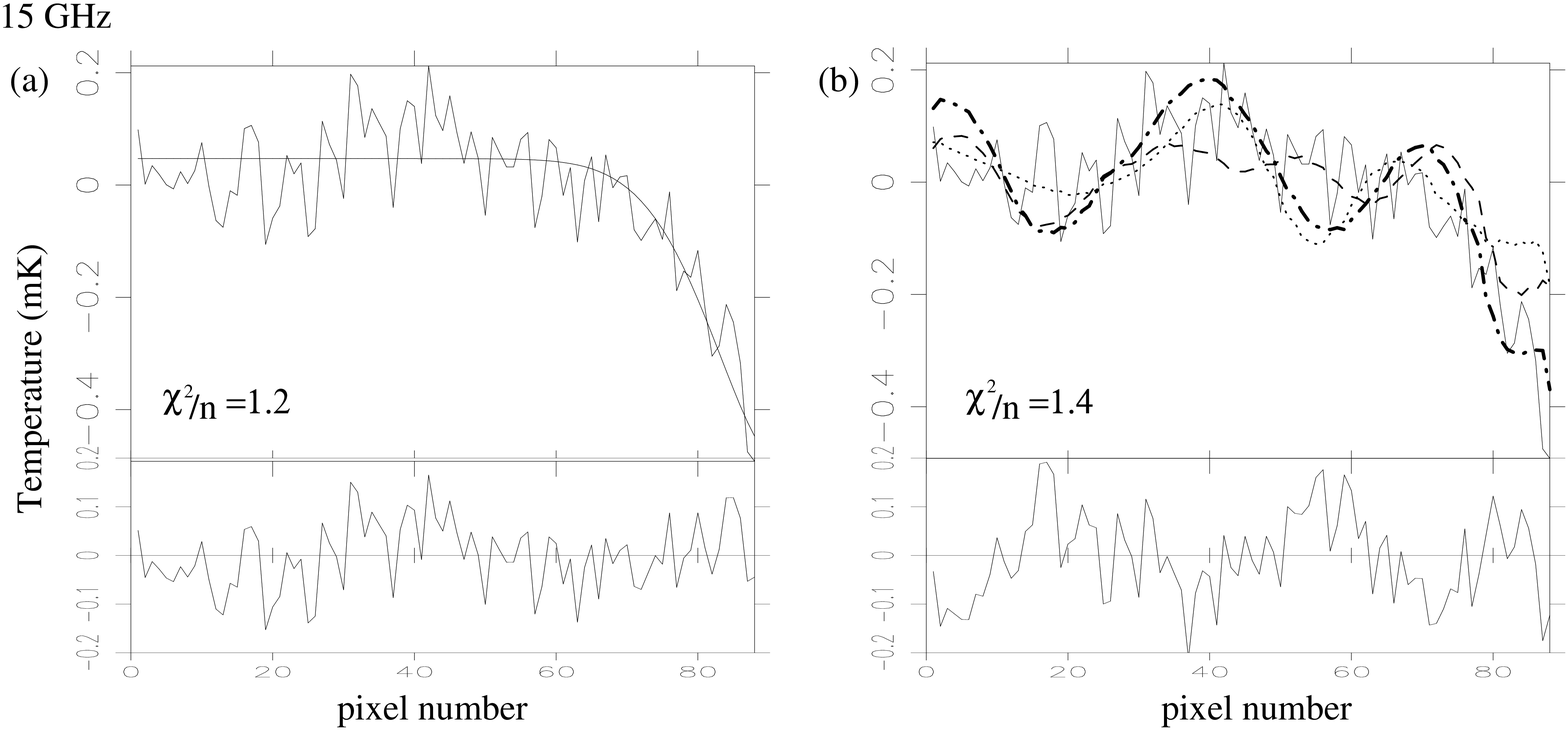,height=4.5cm,width=12cm,angle=0}}
\caption{The Gaussian model (a) and the joint fit result (b) 
for the 15 GHz data at Dec 42.5} 
\end{figure*}

A significant correlation between the Tenerife data and the
DIRBE dust template only occurs close to the Galactic plane where
there is a rise in the overall Galactic signal. This rise may be frequency
dependent if different physical components are present (for example
free-free and synchrotron) so that care must be taken when comparing 
correlation results between different frequencies. 
We take a simple model of the Galaxy 
$$S(b)=S_\circ \exp\left(-{b^2 4 \ln2 \over FWHM^2}\right)$$
where $b$ is the Galactic latitude, $S$ is the amplitude and $FWHM$ is
the full width half maximum of the Gaussian model for the Galactic plane. 
The value of $S_\circ$ is arbitrary as 
it will only scale the estimate for $a$ for the various correlation
results and not affect the relative value between models with different 
width.  Each model is then convolved with the appropriate
experimental beam, and correlated with the data, while varying FWHM. 

Figure 4 shows the widths of the Galactic models which correlated best with 
the data and the Galactic templates for the two declination strips 
in the centre region which do show convincing correlation. Two strips 
showed no significant correlation between the data and the templates, 
and Dec $= 42.5\deg$ has the strong feature at 10~GHz, which is also 
incompatible with the templates. It can be seen that the most significant
correlation between the dust/Haslam template and the model occurs with 
Galactic models of different sizes (FWHM), which are also different 
from the sizes obtained for the 10 and 15GHz data. However there seems to be a 
clear trend of decreasing width with increasing frequency compatible 
with more synchrotron emission at 10~GHz and more dust-correlated 
structure at 15~GHz. The combined fit to all declination stripes 
simultaneously also shows this expected trend of increasing Galactic 
width with wavelength. If this systematic change is real we might 
be able to fit for two distinct components with different width and 
spectral index. Although our fits of a single Gaussian component are 
all well acceptable the quality of the data is however not good enough 
to fit for two such components, if they are 
present independently of templates.

In Figure 4 the points for the dust correlations lie closer to the 
15~GHz points than to the 10~GHz points, implying a better match of 
the Galactic shape for 15 GHz. When correlating the dust template to 
the data, the significance of any correlation will drop and $\hat{a}$ will 
change systematically, because the structure at each frequency is 
different. When correcting for this mismatch, assuming that the correlated 
component we are looking for has identical appearance at any frequency, 
we find that the correlation amplitudes for both 10 and 15 GHz shift 
upwards. However the 10 GHz amplitude, because of the larger difference 
in size with dust, gets a larger shift, which results in a drop of the 
spectral index for the dust-correlated emission by typically about 0.2. 
Taking this effect of mismatched structure into account for the Haslam 
correlations results in a less steep spectral index, which is more 
consistent with synchrotron emission.

When we subtract the best fitting Galactic models from the data and the 
templates no correlations remain between the residuals, neither between 
the data and the templates nor between the templates, showing again that 
any correlation is only due to the Galactic plane signal, which we are 
able to model. Figure 5 shows these best fit models for the 10 and 15GHz 
data, for Dec 42.5, and the residuals.  The fit achieved by the single 
component Gaussian model is indeed good and in fact better than the joint 
correlation fit, which we show for comparison.  Further we find that since 
the joint fit to the 10 GHz data is rather poor in terms of goodness of fit, 
synchrotron and free-free components with the expected spectral indices
 (-2.8 and -2.1) can be 
fitted with an almost similar goodness of fit, based on the reduced $\chi^2$. 
When the same is done for the other declinations the goodness of fit 
is generally better for both the joint fit as well as the model fitting. 
Further in the other declinations also we generally find equally 
acceptable fits for a dust-correlated component which is constrained 
to give a free-free spectral index. 

%\onecolumn  		

\section{Template errors}

Another yet unaccounted source of error is an intrinsic error in the 
dust template itself, which can be instrument noise or data reduction 
errors due to point source subtraction etc., but also systematic variations 
in the dust emissivity, which might not be followed in the same way by the 
components at 10 and 15~GHz. Note that the DIRBE maps at different frequencies 
are correlated with each other to not more than $95 \%$. If we assume that 
there is a $5\%$ Gaussian error in the flux at $100~\mu$m due to these errors 
we can model this effect to quantify any change that would occur in the 
correlation. To do this we added a $5\%$ error in the flux (on the Tenerife 
angular scales) to the maps and calculated the new values of $a$. 
We performed 300 Monte-Carlo simulations and found that there was 
a systematic drop in $a$ for the 10 and 15~GHz data by factors of 
$1.6$ and $1.4$ respectively, when the centre regions of all declinations 
were taken together for the $b>20\deg$ cut (similar values are obtained when 
the entire $b>20\deg$ region is taken). Turning this effect around for
 the real data, assuming 
that there is an intrinsic $5\%$ error present in the template map, we need 
to increase the values of $a$ given in the tables by these factors.  The 
value of $a$ at 10~GHz has to be increased relative to 15~GHz, which results 
in a significantly steeper spectral index, again more consistent with 
free-free emission.  However, it should be noted  
that we do not have error maps for the templates at present and thus in this current paper we have not attempted a quantitative evaluation of this effect
Presumably template errors, given that we have an
 idea of the likely causes, will be greater towards the Galactic plane.  
Since we have found that the only correlations we have to report result for 
pixels close to the Galactic centre, this will be an important source of 
systematic error.

\section{Conclusions} 

Spinning dust emission can be identified and discriminated from free-free 
emission {\em if} indications for a peak in the emission spectrum can be found 
at low microwave frequencies, which are probed by the Tenerife experiment. 
A rising spectrum between 10 and 15~GHz can be taken as a prediction of 
the spinning dust hypothesis, although the exact location of the emission 
peak depends on details of the model. Tentative evidence for a turnover in the 
spectrum of the Galactic dust-correlated  
microwave component between 10 and 15~GHz has been presented by DOC99. 
We however find that the spectral index of dust-correlated emission is 
negative for all Galactic cuts 
except for the $b > 20\deg$ cut. The Galactic signal, and with it the 
significance of the correlation, decreases with increasing Galactic latitude, 
and no correlations are detected in the higher Galactic latitude regions of 
$b > 30\deg$. The variance in the value of $\hat{a}$ between 
regions with different Galactic cuts is rather large.
We further find that the correlation detected in the $b> 20\deg$ region
comes only from a small number of pixels at low Galactic latitude and 
towards the Galactic centre, where signal from the Galactic plane is 
present. This correlation shows a free-free like spectral index, whereas 
the rest of the region was found to be uncorrelated, or even significantly 
anticorrelated. 

%A systematic effect arises due to the presence of significant correlation 
%between the dust and Haslam templates close to the Galactic plane.
Using sky rotations we show that the correlation we see at $b > 20\deg$ 
is only an alignment of structure due to the rise of the Galactic plane 
signal. Employing a simple model for this structure we were able to 
demonstrate that the spatial distribution of Galactic emission is in fact 
different in the templates and in the data, giving rise to a systematic 
error. We were also able to show that 
this simple model of the Galaxy fits the data generally 
better than the templates.  
Further, modelling a Galactic free-free component, which correlates with the 
dust template, generally yields an equally acceptable fit to the data. Another 
significant systematic effect arises due to intrinsic errors in the 
templates, and all these effects cause a misleading increase in the 
inferred spectral index of the dust-correlated component between 10 and 
15~GHz. 

A comparison of our results with 
other experiments is presented in Figure 7.  Here the data points for the 
Tenerife experiment, obtained by taking a weighted 
average of all detections (all regions of all Galactic cuts, joint analyses)
 with errors taken from the sky rotations, correspond
 to values of $180\pm47$ and $123\pm16$ $\mu$K /MJy sr$^{-1}$ at 10 and 15~GHz 
respectively.
Note that the 10GHz point on the plot is significantly higher, as compared to 
that plotted in DOC99 and the value quoted in our table 2
This is because this region consists 
of parts that are correlated as well as parts that do not correlate or even 
anticorrelate, as in the case of the 10GHz data. Here, since we are 
 focussing only on the regions that correlate (these regions are the same for 
both 10 and 15~GHz data and have been found 
to lie close to the Galactic plane), the value is significantly higher. 
Note that the high values of $a$ that we get need not be in 
contradiction to the typical level of free-free allowed by 
$H_{\alpha}$ maps of other regions as our 
detections are close to the Galactic plane, where the level of Galactic 
emission is expected to be high. 

\begin{figure}
\centerline{\epsfig{
file=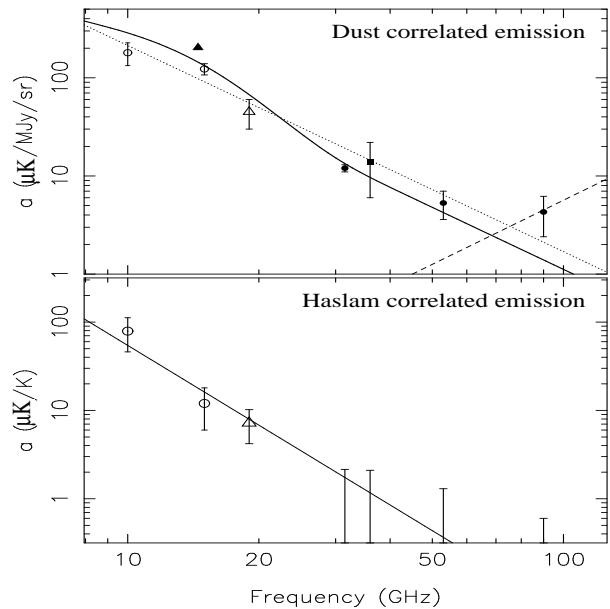,height=8cm,width=8cm,angle=0}}
\caption{Summary of our results on the dust-correlated (top panel) and 
synchrotron-correlated (bottom panel) emission compared to results from other 
experiments (COBE DMR-filled circles, 19GHz-empty triangle, OVRO-filled 
triangle, Saskatoon-filled square). The straight lines 
represent synchrotron (solid), free-free (dotted) and vibrational dust 
(dashed) spectral indices. The thick solid curve shows a combination of 
free-free emission with a small spinning dust contribution with a peak at 
15 GHz in an attempt to fit all 
the data. This fit appears acceptable, as does the single free-free 
component fit, but it should be kept in mind that the data from different 
experiments might not be directly comparable, since they were taken on 
different angular scales and towards different regions of the sky. The 
Tenerife data points are an average derived from our various results, see text
 for details.} 
\end{figure}

The spectral index for dust-correlated emission as 
deduced from the 10 and 15 GHz points is less negative (by about $1\sigma$) 
than expected for free-free emission. 
The spectral index deduced simultaneously from 
synchrotron-correlated emission is systematically steeper than expected. 
This could be attributed 
to the effects that we have identified 
 and which have all been shown to influence the correlation in the 
same direction.  Note also that in this plot only the Tenerife and 
COBE DMR points, at different frequencies, each represent data which 
probe the same angular scales and were taken with the same sky coverage. 
Inferring a spectral index by comparing different experiments assumes 
that the Galactic component is traced by a given fixed template and 
does not depend on parameters which vary over the sky, but our analysis 
of the Tenerife data shows large variations of Galactic correlation 
with the sky region. 

With the present analysis of Tenerife data we are not 
able to make a firm claim about the origin of the dust-correlated component, 
since we do not find convincing support for a spinning dust component.  
This does not have to rule out this hypothesis, since 
environmental conditions or grain sizes, 
which affect the position of the 
spectral peak, could change systematically 
with location on the sky, particularly in the transition region between low 
and high Galactic latitude.  The spatial variation in the correlation 
amplitude and the possibility of the presence of some spinning dust emission 
along with free-free emission need to be dealt with.  
The separation task is difficult for two reasons.  Firstly neither 
the frequency dependence nor the spatial distribution of any of the Galactic 
components at low microwave frequencies is particularly well known. And further
 we would expect these components to be correlated, since we find their
 templates to be correlated, at least at low Galactic latitudes.
The combination of the results from the other experiments, shown in Figure 7, 
might not be strongly constraining, but nevertheless, does 
not give conclusive evidence for spinning dust emission either, without 
an expectation of the amplitude of free-free emission based on dust-correlated 
H$_\alpha$ emission. 

In order to make more reliable inferences, a pixel by pixel separation
of components would have to be performed, using the Maximum Entropy
method for example.  In a forthcoming paper we shall perform such a
separation incorporating information about spinning dust. 
Also, including other data at frequencies lower than 10~GHz and adding
other templates of Galactic emission such as the $H_\alpha$ maps from
the Wisconsin H-Alpha mapper (Tufte, Reynolds and Haffner 1998)
would be useful.

\section*{Acknowledgements}  
\label{lastpage}

We wish to thank all the people involved in taking the Tenerife
data set, and Juan Francisco Macias-Perez for useful comments.
 PM acknowledges financial support from the Cambridge Commonwealth Trust. 
AWJ acknowledges King's College, Cambridge for support in the
form of a Research Fellowship. RK acknowledges support from an EU
Marie Curie Fellowship.

%\onecolumn
%\begin{table}
%\begin{center}
% {\small
%\caption{The following table shows the correlation coefficients $\hat{a}\pm \Delta(\hat{a})$ obtained when only the Galactic centre half of the data at Dec $42.5$ was correlated with dust, and when the Galactic centre half of all the declinations together was correlated with dust.  At Dec $42.5$ there seems to be a strong synchrotron feature.  Also note that the value of $\hat{a}$ changes lot between declinations.  Though all the declinations are not shown here, the variance in $\hat{a}$, for correlation with dust, between declinations is large. See text.}
%\label{tab:table_zero} 
%\begin{tabular}{|c|c|c|c|c|c|} \hline
%\multicolumn{2}{c}{} & \multicolumn{2}{c}{10GHz} &
%\multicolumn{2}{c}{15GHz} \\ \cline{3-6}
%region  &  & $with dust$ & $with haslam$ & $with dust$ & $with haslam$\\ \hline
%DEC 42 & individual & $1016\pm 72$ & $382\pm 20$ & $437\pm 36$ & $143\pm 12$\\[-0mm]
%centre & joint & $225\pm 95$ & $341\pm 27$ & $261\pm 55$ & $78\pm 18$ \\ \hline
%all DECs & individual & $307\pm 29$ & $110\pm 8$ & $213\pm 18$ & $47\pm 6$\\[-0mm]
%centre & joint & $160\pm 33$ & $89\pm 9$ & $182\pm 22$ & $27\pm 7$ \\ \hline
%\end{tabular}
%}
%\end{center}
%\end{table}

\onecolumn
\begin{table}
\begin{center}
 {\small
\caption{Individual correlations for the Tenerife data.  $\hat{a}$ is the
correlation coefficient, and has units $\mu K(MJy/sr)^{-1}$ for the
$100\mu$m template, and $\mu K/K$ for the Haslam and $R\&R$ templates.}
\label{tab:table_one} 
\begin{tabular}{|c|c|c|c|c|c|} \hline
\multicolumn{2}{c}{} & \multicolumn{2}{c}{10GHz} &
\multicolumn{2}{c}{15GHz} \\ \cline{3-6}
b  & Template & $\hat{a}\pm \Delta(\hat{a})$ & $\Delta T \mu K$ & $\hat{a}\pm \Delta(\hat{a})$
& $\Delta T \mu K$ \\ \hline
$b > 0\deg$ & $100\mu m$ & $214\pm 0$ & $6435\pm 13$ & $119\pm 0$ & $3430\pm 6$ \\[-0mm]
    & Has & $591\pm 1$ & $5850\pm 12$ & $333\pm 1$ & $3189\pm 6$ \\[-0mm]
    & $R\&R$ & $26140\pm 64$ & $4684\pm 11$ & $14176\pm 33$ & $2476\pm
5$ \\ \hline

$b > 5\deg$ & $100\mu m$ & $209\pm 1$ & $3127\pm 10$  & $123\pm 0$ &
    $1770\pm 5$ \\
    & Has & $545\pm 2$ & $2617\pm 9$ & $323\pm 1$ & $1502 \pm 4$ \\
    & $R\&R$  & $17324\pm 86$ & $1773\pm 9$ & $10790\pm 41$ & $1077\pm 4$ \\ \hline

$b > 10\deg$ & $100\mu m$ & $268\pm 1$ & $1918\pm 11$ & $136\pm 1$ &  $967\pm 4$   \\[-0mm]
    & Has & $431\pm 3$ & $1422\pm 9$& $259\pm 1$ & $839\pm 4$  \\[-0mm]
    & $R\&R$  & $13209 \pm 98$ & $1079\pm 8$& $8654\pm 47$ & $692\pm 4$ \\ \hline

$b > 15\deg$ & $100\mu m$ & $259\pm 9$ & $199\pm 7$& $132\pm 4$ &$106\pm 3$  \\
    & Has & $142\pm 5$ & $161\pm 6$ & $81\pm 3$ &$90\pm 3$  \\ 
    & $R\&R$  & $4727\pm 148$ & $207\pm 6$& $2883\pm 84$ &  $125\pm 4$\\ \hline

$b > 20\deg$ & $100 \mu m$ & $131\pm 21$ & $41\pm 7$& $140\pm 13$ &  $43\pm 4$  \\[-0mm]
    & Has & $71\pm 6$ & $70\pm 6$& $29\pm 4$ &  $28\pm 4$\\[-0mm]
    & $R\&R$ & $2294\pm 256$ & $54\pm 6$ & $1514\pm 165$ &$35\pm 4$\\ \hline

$b > 30\deg$ & $100 \mu m$ & $18\pm 38$ & $4\pm 8$& $32\pm 26$ &$6\pm 5$ \\[-0mm]
    & Has & $20\pm 8$ & $20\pm 8$ & $7\pm 4$ &$6\pm 4$\\[-0mm]
    & $R\&R$ & $897\pm 390$ & $17\pm 7$ & $379\pm 255$ &$7\pm 5$\\ \hline

$b > 40\deg$ & $100 \mu m$ & $39\pm 43$ & $8\pm 8$ & $13\pm 30$ &$2\pm 6$ \\
    & Has & $18\pm 10$ & $14\pm 8$ & $10\pm 7$ &$8\pm 5$\\[-0mm]
    & $R\&R$ & $964\pm 432$ & $18\pm 8$ & $531\pm 304$ &$10\pm 5$\\ \hline
\end{tabular}
}
\end{center}
\end{table}

\begin{table}
\begin{center}
{\small
\caption{Joint correlations for the Tenerife data.  $\hat{a}$ is the
correlation coefficient, and has units $\mu K(MJy/sr)^{-1}$ for the
$100\mu$m template, and $\mu K/K$ for the Haslam and $R\&R$ templates.}
\label{tab:table_three} 
\begin{tabular}{|c|c|c|c|c|c|} \hline
\multicolumn{2}{c}{} & \multicolumn{2}{c}{10GHz} &
\multicolumn{2}{c}{15GHz} \\ \cline{3-6}
b  & Template & $\hat{a}\pm \Delta(\hat{a})$ & $\Delta T \mu K$& $\hat{a}\pm \Delta(\hat{a})$ & $\Delta T \mu K$\\ \hline
$b > 0\deg$ & $100\mu m$ & $151\pm 1$ & $4535\pm 36$ & $114\pm 1$ &$3281\pm 19$\\
   & Has & $191\pm 3$ & $1895\pm 33$ & $16\pm 2$ &$157\pm 18$  \\
   & $R\&R$  & $6829\pm 90$ & $1224\pm 16$ & $1645\pm 52$ &$287\pm 9$\\ \hline

$b > 5\deg$ & $100\mu m$ & $154\pm 1$ & $2311\pm 21$ & $116\pm 1$ &$1669\pm 10$\\
   & Has & $171\pm 4$ & $821\pm 19$ & $23\pm 2$ &$106\pm 10$\\
   & $R\&R$  & $5587\pm 99$ & $572\pm 10$ & $1405\pm 54$ &$140\pm 5$\\ \hline

$b > 10\deg$ & $100\mu m$ & $238\pm 3$ & $1704\pm 20$ & $136\pm 1$ &$961\pm 10$\\
   & Has & $64\pm 5$ & $213\pm 17$& $0\pm 3$ &$0\pm 10$  \\
   & $R\&R$  & $1701\pm 140$ & $139\pm 11$ & $394\pm 79$ &$31\pm 6$\\ \hline

$b > 15\deg$ & $100\mu m$ & $200\pm 10$ & $154\pm 7$ & $110\pm 5$ &$88\pm 4$\\
   & Has & $94\pm 6$ & $106\pm 7$ & $26\pm 4$ &$29\pm 4$\\
   & $R\&R$  & $3393\pm 211$ & $148\pm 9$ & $1858\pm 140$ &$81\pm 6$\\ \hline

$b > 20\deg$ & $100\mu m$ & $67\pm 22$ & $21\pm 7$ & $122\pm 14$ &$38\pm 4$\\
   & Has & $65\pm 7$ & $64\pm 6$ & $12\pm 4$ &$11\pm 4$\\
   & $R\&R$  & $1983\pm 288$ & $47\pm 7$& $726\pm 203$ & $17\pm 5$\\ \hline

$b > 30\deg$ & $100\mu m$ & $-16\pm 40$ & $-3\pm 8$& $18\pm 28$ &$4\pm 6$ \\
   & Has & $21\pm 8$ & $19\pm 7$ & $6\pm 5$ &$5\pm 5$\\
   & $R\&R$  & $1087\pm 455$ & $20\pm 8$ & $291\pm304$ &$5\pm 5$\\ \hline

$b > 40\deg$ & $100\mu m$ & $0\pm 49$ & $0\pm 10$ & $-21\pm 37$ &$-4\pm 7$\\
   & Has & $18\pm 11$ & $14\pm 9$ & $13\pm 8$ &$10\pm 6$\\
   & $R\&R$  & $1245\pm 576$ & $23\pm 10$ & $842\pm 421$ &$15\pm 8$\\ \hline
\end{tabular}
}
\end{center}
\end{table}

\begin{table}
\begin{center}
{\small
\caption{Spectral indices obtained from individual and joint correlations. No 
number is given when the correlation is insignificant.}
\label{tab:table_five} 
\begin{tabular}{|c|c|c|c|c|} \hline
\multicolumn{1}{c}{} & \multicolumn{2}{c}{Individual analysis} &
\multicolumn{1}{c}{joint analysis} \\ \cline{2-5}
cut & dust & Haslam & dust & Haslam \\ \hline
$b > 0\deg$ & $-1.45^{+0.01}_{-0.01}$ & $-1.41^{+0.01}_{-0.01}$ & $-0.69^{+0.04}_{-0.04}$ & $-6.11^{+0.34}_{-0.37}$ \\ [2mm]
$b > 5\deg$ & $-1.31^{+0.01}_{-0.01}$ & $-1.29^{+0.02}_{-0.02}$ & $-0.70^{+0.04}_{-0.04}$ & $-4.95^{+0.27}_{-0.28}$ \\ [2mm]
$b > 10\deg$ & $-1.67^{+0.03}_{-0.03}$ & $-1.25^{+0.02}_{-0.03}$ & $-1.38^{+0.05}_{-0.05}$ & $-$  \\ [2mm]
$b > 15\deg$ & $-1.66^{+0.16}_{-0.16}$ & $-1.38^{+0.17}_{-0.18}$ & $-1.47^{+0.23}_{-0.24}$ & $-3.17^{+0.52}_{-0.56}$ \\ [2mm]
$b > 20\deg$ & $0.16^{+0.65}_{-0.60}$ & $-2.21^{+0.54}_{-0.56}$ & $1.48^{+1.35}_{-1.00}$ & $-4.17^{+1.00}_{-1.25}$ \\ [2mm]
$b > 30\deg$ & $-$ & $-2.56^{+2.35}_{-2.95}$ & $-$ & $-3.09^{+2.68}_{-5.21}$ \\ [2mm]
$b > 40\deg$ & $-$ & $-1.45^{+3.31}_{-4.06}$ & $-$ & $-0.80^{+3.51}_{-3.53}$ \\ \hline
\end{tabular}
}
\end{center}
\end{table}

\begin{table}
\begin{center}
 {\small
\caption{Joint correlations obtained for data at different declinations when the data at each declination are divided into two halves, one half being towards the Galactic centre (marked C) and the other towards the anticentre (marked AC).}
\label{tab:table_zero} 
\begin{tabular}{|c|c|c|c|c|c|c|c|} \hline
\multicolumn{2}{c}{} & \multicolumn{3}{c}{10GHz} &
\multicolumn{2}{c}{15GHz} \\ \cline{3-8}
region  & number of pixels &  $dust$ & $Haslam$ & $\chi^2$ & $dust$ & $Haslam$ & $\chi^2$\\ \hline
all Decs, C & 426 & $160\pm 33$ & $89\pm 9$ & $856$ & $182\pm 22$ & $27\pm 7$ & $530$ \\
all Decs, AC & 424 & $-80\pm 32$ & $15\pm 10$ & $481$ & $43\pm 21$ & $-50\pm 6$ & $469$ \\ \hline
Dec 42.5, C & 88 & $225\pm 95$ & $341\pm27$ & $259$ & $261\pm 55$ & $78\pm 18$ & $122$ \\
Dec 42.5, AC & 87 & $-185\pm 73$ & $29\pm 19$ & $107$ & $24\pm 43$ & $-12\pm 12$ & $88$ \\ \hline
Dec 40.0, C & 86 & $276\pm 56$ & $74\pm 33$ & $94$ & $225\pm 41$ & $83\pm 22$ & $93$ \\
Dec 40.0, AC & 86 & $-148\pm 55$ & $-12\pm 18$ & $94$ & $21\pm 34$ & $-1\pm 12$ & $79$ \\ \hline
Dec 37.5, C & 85 & $177\pm 50$ & $89\pm 22$ & $100$ & $173\pm 32$ & $25\pm 14$ & $69$ \\
Dec 37.5, AC & 85 & $36\pm 52$ & $37\pm 17$ & $100$ & $17\pm 32$ & $-5\pm 10$ & $79$ \\ \hline
Dec 35.0, C & 84 & $75\pm 74$ & $50\pm 13$ & $85$ & $72\pm 45$ & $12\pm 10$ & $86$ \\
Dec 35.0, AC & 84 & $0\pm 54$ & $-8\pm 21$ & $86$ & $17\pm 33$ & $-13\pm 10$ & $106$ \\ \hline
Dec 32.5, C & 83 & $215\pm 94$ & $-12\pm 20$ & $57$ & $-77\pm 69$ & $38\pm 15$ &$71$ \\
Dec 32.5, AC & 82 & $-75\pm 92$ & $7\pm 24$ & $79$ & $128\pm 43$ & $13\pm 11$ & $89$ \\ \hline    
\end{tabular}
}
\end{center}
\end{table}

\begin{table}
\begin{center}
 {\small
\caption{The following table shows the correlation coefficients $\hat{a}\pm \Delta(\hat{a})$ obtained when the 20 pixels with lowest Galactic latitude in the Galactic centre region in each declination are correlated with the corresponding structure in the template maps, and those obtained when the remaining pixels are correlated.  All these are joint fits. }
\label{tab:table_zero} 
\begin{tabular}{|c|c|c|c|c|c|c|c|} \hline
\multicolumn{2}{c}{} & \multicolumn{3}{c}{10GHz} &
\multicolumn{2}{c}{15GHz} \\ \cline{3-8}
region  & number of pixels &  $dust$ & $Haslam$ & $\chi^2$ & $dust$ & $Haslam$ & $\chi^2$\\ \hline
all Decs & 100 & $178\pm 44$ & $111\pm 14$ & $479$ & $167\pm 31$ & $18\pm 11$ & $159$ \\
all Decs & 750 & $-52\pm 27$ & $22\pm 8$ & $821$ & $29\pm 19$ & $47\pm 51$ & $808$ \\ \hline
Dec 42.5 & 20 & $1148\pm 374$ & $279\pm 49$ & $40$ & $466\pm 141$ & $24\pm 29$ & $24$ \\
Dec 42.5 & 155 & $-147\pm 57$ & $37\pm 17$ & $187$ & $54\pm 35$ & $-14\pm 11$ & $174$\\ \hline
Dec 40.0 & 20 & $264\pm 90$ & $48\pm 10$ & $26$ & $177\pm 59$ & $35\pm 65$ & $24$ \\
Dec 40.0 & 152 & $-75\pm 47$ & $0\pm 16$ & $160$ & $26\pm 30$ & $10\pm 10$ & $139$ \\ \hline
Dec 37.5 & 20 & $315\pm 144$ & $163\pm 92$ & $19$ & $217\pm 75$ & $64\pm 54$ & $10$ \\
Dec 37.5 & 150 & $10\pm 46$ & $51\pm 15$ & $180$ & $6\pm 29$ & $1\pm 8$ & $129$ \\ \hline
Dec 35.0 & 20 & $23\pm 121$ & $53\pm 21$ & $20$ & $34\pm 77$ & $3\pm 18$ & $23$ \\
Dec 35.0 & 148 & $21\pm 48$ & $14\pm 15$ & $150$ & $7\pm 30$ & $-6\pm 8$ & $170$ \\ \hline
Dec 32.5 & 20 & $152\pm 200$ & $9\pm 58$ & $18$ & $32\pm 152$ & $8\pm 42$ & $10$ \\
Dec 32.5 & 145 & $34\pm 73$ & $0\pm 17$ & $125$ & $79\pm 39$ & $18\pm 9$ & $155$ \\ \hline
\end{tabular}
}
\end{center}
\end{table}

%\label{lastpage}

\end{document}